\documentclass[showpacs,preprintnumbers,amsmath,amssymb]{revtex4-2}
\usepackage[utf8]{inputenc}
\usepackage[english]{babel}
\usepackage{graphicx}
\usepackage{amsfonts}
\begin{document}
\title{Is bouncing easier with a negative effective dark fluid density ?}
\author{Stéphane Fay\footnote{steph.fay@gmail.com}}
\affiliation{
Palais de la Découverte\\
Astronomy Department\\
Avenue Franklin Roosevelt\\
75008 Paris\\
France}
\begin{abstract}
Assuming that a cosmological model can describe the \emph{whole} Universe history, we look for the conditions of a cosmological bounce that thus have to be in agreement with late time observations. Our approach involves casting such a theory into General Relativity with curvature ($\Omega_{\kappa}$), matter ($\Omega_{m}$), radiation ($\Omega_{r}$) and an effective dark fluid ($\Omega_{d}$) and formulating the corresponding field equations as a 2D dynamical system, wherein phase space points corresponding to extrema of the metric function are constrained by observational data on the aforementioned parameters densities. We show that if this effective dark fluid density is positive at the bounce, these observational constraints imply its occurrence in the future at a redshift $z<-0.81$ whatever the cosmological model (dark energy, brane, $f(R)$, etc.) corresponding to this effective dark fluid and even in the presence of positive curvature. Hence, the effective dark fluid density must be negative at the bounce such as it arises for $z>-0.81$ and thus possibly in the past. Observations also impose that the dark fluid effective density can change sign only within the redshift range $0.54<z<0.61$. We then proceed by examining three distinct cosmological models: a non linear dark fluid model, a Randall Sundrum brane model and a $f(R)=R+mR^n+\Lambda$ model. For each of them, we examine the conditions for (1) a bounce at early time (when radiation dominates over matter), (2) with a negative effective dark fluid energy density, (3) this having to change sign within the above specified redshift interval (to be positive at present time). We find that none of the three models under consideration satisfy all three constraints. We therefore conclude that while a negative effective dark fluid energy density required by observational constraints on $\Omega_m$, $\Omega_r$ and $\Omega_{k}$ for a bounce at early times facilitates this bounce, the requirement for $\Omega_d$ to change sign and become positive within the above specified narrow redshift interval proves exceedingly challenging to satisfy these same constraints.
\end{abstract}
\maketitle
\section{Introduction} \label{s0}
In this paper, we investigate how a classical cosmological bounce can be constrained by late time observations and how it is difficult for it to satisfy them. A cosmological bounce is a way to circumvent a singularity. With an ekpyrotic scenario, it can also produce effects similar to inflation\cite{Gut81,Lin08} such as smoothing out the cosmological background\cite{Kho01,PetPin08,QuiBra14,BraPet17,BraWan20}. Two types of bounce can be distinguished: a quantum one \cite{GieTur16, Pin21} or a classical one \cite{BatPet14}. Loop Quantum Cosmology \cite{Ash09,Agu16} is the most successful quantum bounce in which the singularity is resolved in a generic way\cite{AshPaw06}. It predicts a quantum bridge between contracting and expanding Universe. In the classical scenario, after contracting to a finite size larger than the Planck length, the Universe undergoes a bounce and begins expanding. This is generally challenging since such bounce has to violate the null energy condition, leading to instability and ghost that can nerveless be avoided \cite{KumMah21, IjjSte16, BisGer12}. In this paper, we focus on a classical bounce and consider a cosmological model that includes matter, radiation, (effective or not) dark fluid and curvature, encompassing the \emph{whole} Universe history from early to late times\cite{PerLim13,LimBas13,LimPer15}. We show then how late time observations imply that the effective dark fluid density must be negative during the bounce and change sign within a narrow redshift interval. By examining several cosmological models (a non linear dark fluid model, a Randall Sundrum brane model and a $f(R)=R+mR^n+\Lambda$ model) for which we calculate the corresponding effective dark fluid equation of state $w$, we show how it is challenging to achieve such a bounce consistent with current observations of parameters density. Let us note that curvature is often neglected whereas it provides insights into various early Universe scenarios: emergence from a false vacuum\cite{ColLuc80, BouHar15, RafBir15}, multiverse \cite{Teg03, CesAlw20, FreKle06} or initial conditions\cite{Tak15}. Methods for detecting curvature have been developed\cite{Ber06,Den18} but observational evidence is still pending \cite{Hin13, Val20, Han21} while some theories suggest that it should be either flat or closed\cite{ValAnc21, EllMcE02}.\\
Studying a bouncing Universe benefits from employing mathematical tools that provide insight into the general properties of an equation system, such as extrema of the metric function, without necessarily solving it. Dynamical system analysis\cite{Wai97, Boh17, Bah18, Leo24, Kad23, Ros24} serves as an effective approach for this purpose. By reformulating the field equations into a dynamical system represented by trajectories in a phase space, we transform the task of studying a system of differential equations into one of investigating the geometric properties of this phase space. However, each gravitation theory needs some specific variables to be rewritten as a dynamical system whereas they can generally be cast into General Relativity with an effective dark fluid by rearranging the terms in the field equations (i.e. without a conformal transformation of the metric) leaving unchanged the scale factor evolution and thus the existence of a bounce. Hence, as a mathematical tool, we will define a phase space for General Relativity with matter, radiation, an effective dark fluid and curvature. Then we will use this tool to study bouncing cosmology for any theories, from General Relativity to $f(R)$ theories, defining their corresponding effective dark fluid. Finding some variables describing such a phase space in presence of curvature is a difficult task (see \cite{Bah18,Ker19, Ker20} for instance). We manage to use only two normalised variables related to the matter and radiation densities. The dynamical system thus consists in a set of only two differential equations, whereby the trajectories in phase space depend on the effective dark fluid equation of state $w$ and a constant $p$\\
From this phase space we get the following results. First, observations\cite{Agh20} constrain $0<p<0.27\times 10^{-4}$. This inequality determines the phase space points where extrema of the metric function occur and thus a bounce when also specifying $w$. Furthermore, it also allows to derive that a positive effective dark fluid density at the bounce invariably entails its occurrence in the future at a redshift $z<-0.81$, whatever the cosmological model describing the whole Universe history. Hence, a negative effective dark fluid density is necessary for a bounce in the past that agrees with the present day observations, even in the presence of a positively curved Universe. Moreover, observations pinpoint the redshift range within which the effective dark fluid density undergoes a sign change, namely $0.54<z<0.81$, this inequality being also model-independent. We then examine the above mentioned phase space for three cosmological models: a nonlinear dark fluid model, a Randall Sundrum brane model and a $f(R)=R+mR^n+\Lambda$ model. For each of them, we determine whether the Universe can undergo a bounce at early time, when radiation dominates matter, with a negative effective (or not for the first model) dark fluid energy density capable of transitioning to a positive value within the above specified redshift range: none of the models fulfils all these constraints. We then conclude negatively to the title of this paper: while a bounce is facilitated by a negative effective dark fluid density, the corresponding cosmological model (here dark energy, brane, $f(R)$) typically fails to align with observational data.\\
The plan of the paper is as follows. In section \ref{s1}, we show how to rewrite the field equations of General Relativity with matter, radiation, an effective dark fluid and curvature into a dynamical system with two normalised variables. We explain how to determine the space phase points corresponding to extrema of the metric function and why the effective dark fluid density has to be negative to have a cosmological bounce in the past. In section \ref{s2}, we use the formalism of section \ref{s1} to study three cosmological models: a non linear dark fluid model, a Randall Sundrum brane model and a $f(R)=R+mR^n+\Lambda$ model. We find that none of them can describe a bounce at early time in agreement with present day observations and conclude in the last section. 
\section{Field equations and phase space} \label{s1}
Most of cosmological models can be rewritten as General Relativity with an effective dark fluid in a way preserving the evolution of the scale factor, i.e. by simply rearranging the terms in the field equations. To study three of them in the next section, we thus define a common mathematical framework by casting the equations of General Relativity with matter density $\rho_m$, radiation density $\rho_r$, effective dark fluid density $\rho_d$ and curvature into a finite two dimensional phase space.
\subsection{Dynamical system} \label{s11}
\begin{center}
\begin{figure}[h]
\centering
\includegraphics[width=6cm]{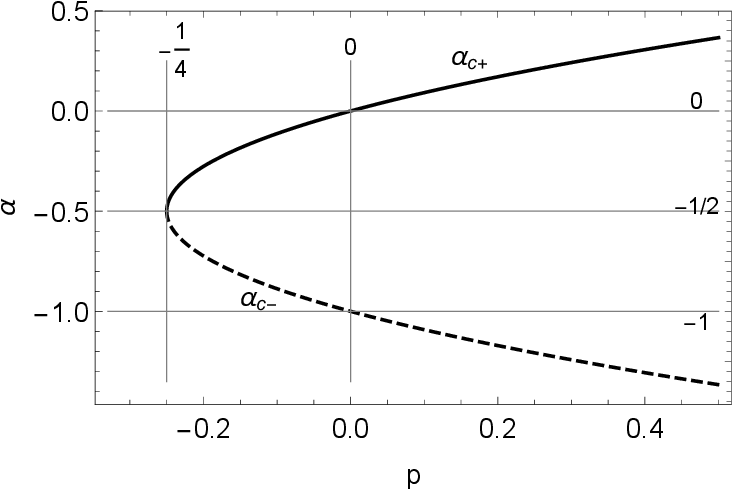}
\caption{\scriptsize{\label{fig4}Values of the contact points $\alpha_{c\pm}$ (respectively plain and dashed curves) between the unit circle $\omega_m^2+\omega_r^2=1$ and the parametric curve (\ref{div}) corresponding to the extrema of the metric function with respect to $p$.}}
\end{figure}
\end{center}

With a FLRW metric, equations of General Relativity with curvature, radiation, matter and an effective dark fluid are
\begin{equation}\label{eqH2}
H^2=\frac{\kappa}{3}(\rho_m+\rho_r+\rho_d)-\frac{k}{a^{2}}
\end{equation}
\begin{equation}\label{eqHp}
\dot H=-\frac{\kappa}{2}(\rho_m+\frac{4}{3}\rho_r+(1+w)\rho_d)+\frac{k}{a^{2}}
\end{equation}
\begin{equation}\label{eqm}
\dot \rho_m+3H\rho_m=0
\end{equation}
\begin{equation}\label{eqr}
\dot\rho_r+4H\rho_r=0
\end{equation}
\begin{equation}\label{eqd}
\dot\rho_d+3(1+w)H\rho_d=0
\end{equation}
"a" is the scale factor of the FLRW metric and $k$ the curvature parameter. It is $0$, negative or positive for respectively a flat, open or closed Universe. A dot means a derivative with respect to the proper time $t$. We define the following variables
\begin{equation}\label{y1}
\Omega_{m}=\frac{\kappa}{3}\frac{\rho_m}{H^2}
\end{equation}
\begin{equation}\label{y2}
\Omega_{r}=\frac{\kappa}{3}\frac{\rho_r}{H^2}
\end{equation}
\begin{equation}
\Omega_{d}=\frac{\kappa}{3}\frac{\rho_d}{H^2}
\end{equation}
\begin{equation}
\Omega_{k}=\frac{k}{a^2 H^2}
\end{equation}
From the energy conservation equations (\ref{eqm}-\ref{eqr}), one gets as usual 
$$
\Omega_m=H_0^2\Omega_{m0}a^{-3}H^{-2}
$$
$$
\Omega_r=H_0^2\Omega_{r0}a^{-4}H^{-2}
$$
where the subscript "0" indicates the values of the quantities at an initial proper time $t_0$ when, without loss of generality and defining $N=\ln a$, $N_0=\ln a_0 = 0$. We then derive 
\begin{equation}\label{metFunc}
a=\frac{\Omega_{r0}\Omega_{m}}{\Omega_{m0}\Omega_{r}}
\end{equation}
\begin{equation}\label{H2pp}
H^2=H_0^2\frac{\Omega_{m0}^4\Omega_{r}^3}{\Omega_{r0}^3\Omega_{m}^4}
\end{equation}
and thus
\begin{equation}\label{curvTerm}
\Omega_{k}=\frac{k}{H_0^2} \frac{\Omega_{r0}\Omega_{m}^2}{\Omega_{m0}^2\Omega_{r}}
\end{equation}
This allows to rewrite the field equations as a dynamical system for the (non normalised) variables $\Omega_m$ and $\Omega_r$, i.e. a system of two first order equations for these variables
$$
\Omega_m'=\Omega_m\left[\Omega_r-3w(\Omega_m+\Omega_r-1)\right]+p(1+3w)\frac{\Omega_m^3}{\Omega_r}
$$
$$
\Omega_r'=\Omega_r\left[\Omega_r-1-3w(\Omega_m+\Omega_r-1)\right]+p(1+3w)\Omega_m^2
$$
A prime stands for a derivative with respect to $N$ and we have the constraint
\begin{equation}\label{cons}
\Omega_m+\Omega_r+\Omega_d-p\frac{\Omega_{m}^2}{\Omega_{r}}=1
\end{equation}
with the constant
\begin{equation}\label{constant}
p=\frac{\Omega_{k_0}\Omega_{r0}}{\Omega_{m0}^2}
\end{equation}
and $\Omega_{{k_0}}= k H_0^{-2}$. The constant $p$ can also be rewritten in a more general way as 
$$
p=\frac{\Omega_{k}\Omega_{r}}{\Omega_{m}^2}
$$
$p$ is thus a constant that can be fixed by some observations at any initial time $N=0$. For a bouncing Universe, the scale factor must have a minimum $a_{min}\not =0$. This occurs when $H=0$, i.e. when $\Omega_m$ and $\Omega_r$ diverge at a finite value $a_{min}$ of the scale factor such as $\Omega_r = \alpha \Omega_m$ with
$$
\alpha=\frac{\Omega_{r0}}{\Omega_{m0}}a_{min}^{-1}
$$
that is thus a constant. To study a cosmological bounce, we need to rewrite the above dynamical system with normalised variables. We choose: 
\begin{equation}\label{var1}
\omega_m=\frac{\Omega_m}{\sqrt{\Omega_m^2+\Omega_r^2+\Omega_d^2}}
\end{equation}
\begin{equation}\label{var2}
\omega_r=\frac{\Omega_r}{\sqrt{\Omega_m^2+\Omega_r^2+\Omega_d^2}}
\end{equation}
Both variables vary in the range $\left[-1,1\right]$ and have respectively the same signs as $\Omega_m$ and $\Omega_r$ (that we do not assume positive in this section to keep it as general as possible). The whole phase space $(\omega_m,\omega_r)$ is thus a unit disk defined by the constraint
$$
\omega_m^2+\omega_r^2 \leq 1
$$
On its circumference, $\Omega_d=0$. With these variables, the field equations take the dynamical system form
\begin{equation}\label{eq1}
\omega_m'=-\omega_m \left[-\omega_r^2+3 w (-1+\omega_m^2+\omega_r^2)\right]
\end{equation}
\begin{equation}\label{eq2}
\omega_r'=-\omega_r \left[1-\omega_r^2+3 w (-1+\omega_m^2+\omega_r^2)\right]
\end{equation}
Its equilibrium points, whose stability can only be determined by specifying the equation of state $w$, are
\begin{itemize}
\item $M_{\pm}:(\omega_m,\omega_r)=(\pm 1,0)$: matter is dominating.
\item $R_\pm: (\omega_m,\omega_r)=(0,\pm 1)$: radiation is dominating.
\item $D: (\omega_m,\omega_r)=(0,0)$: dark fluid is dominating.
\item $DR: (\omega_m,w)=(0,1/3)$: dark fluid plays the role of radiation when radiation and dark fluid dominate matter. In particular in $(\omega_m,\omega_r)=(0,1/\sqrt{2})$, $\Omega_r\rightarrow \Omega_d$ and diverges whereas $\Omega_m/\Omega_r\propto a\rightarrow 0$ and we have a singularity.
\item $DM: (\omega_r,w)=(0,0)$: dark fluid plays the role of matter when matter and dark fluid dominate radiation.
\end{itemize}
Although the constant $p$ does not appear in the dynamical system (\ref{eq1}-\ref{eq2}), its trajectories still depend on the value of $p$: the above dynamical system defines all the possible trajectories whatever $p$ however, depending on the value of $p$, only some parts of these trajectories are physical. Let us show why. This takes roots in the fact that if to $1$ couple of values $(\Omega_m,\Omega_r)$ corresponds $1$ couple of values $(\omega_m,\omega_r)$, the reverse is not true. Hence, reversing definitions (\ref{var1}-\ref{var2}), it comes
$$
(\Omega_r,\Omega_d)=(\frac{\Omega_m}{\omega_m}\omega_r,\pm\frac{\Omega_m}{\omega_m}\sqrt{1-\omega_m^2-\omega_r^2})
$$
or 
\begin{eqnarray}\label{Otoo}
(\Omega_{m},\Omega_{r})=\nonumber\\
(\frac{\omega_m\omega_r^3}{-p\omega_m^2\omega_r^2+\omega_r^3(\omega_m+\omega_r)\pm\sqrt{\omega_r^6(1-\omega_m^2-\omega_r^2)}},\nonumber\\
\frac{\omega_r^4}{-p\omega_m^2\omega_r^2+\omega_r^3(\omega_m+\omega_r)\pm\sqrt{\omega_r^6(1-\omega_m^2-\omega_r^2)}})\\\nonumber
\end{eqnarray}
To one point $(\omega_m,\omega_r)$ thus corresponds two points $(\Omega_m,\Omega_r)$: one with the plus sign in (\ref{Otoo}) when $\Omega_d/\Omega_r>0$ and the other with the minus sign when $\Omega_d/\Omega_r<0$. Then, we have to determine the set of points on the phase space $(\omega_m,\omega_r)$ that corresponds to a set of points on the phase space $(\Omega_m,\Omega_r)$ with the right sign for $\Omega_d/\Omega_r$. Such a set for $(\omega_m,\omega_r)$ have for limits the circumference of the unit circle ($\omega_m^2+\omega_r^2=1$ with $\Omega_d=0$) and the curves in the phase space $(\omega_m,\omega_r)$ where an extremum of the metric function (minimum, maximum, inflexion point) occurs. Since then, as written before, $\Omega_r = \alpha \Omega_m$ with $\Omega_m\rightarrow \pm\infty$ and $\alpha$ a constant, to calculate these curves, we replace $\Omega_r$ by $\alpha\Omega_m$ in the definitions (\ref{var1}-\ref{var2}) for $(\omega_m,\omega_r)$. The limits $\Omega_m \rightarrow \pm \infty$ then define two parametric curves in the plane $(\omega_m,\omega_r)$ on which are located the extrema of the metric function :
\begin{eqnarray}\label{div}
\pm((1+\alpha^2+(1+\alpha-p/\alpha)^2)^{-1/2},\nonumber\\
\alpha(1+\alpha^2+(1+\alpha-p/\alpha)^2)^{-1/2})\\\nonumber
\end{eqnarray}
The "+" and "-" parts of these parametric curves are symmetric with respect to the origin $(\omega_m,\omega_r)=(0,0)$ of the phase space. Therefore, the physically relevant points of the trajectories lie within the circumference of the unit circle and the parametric curves (\ref{div}) and obviously depend on $p$. These parametric curves correspond to the sets of points in the phase space $(\omega_m,\omega_r)$ where the metric function reaches an extremum. These curves are independent of the dark fluid model, unlike the nature of the extremum (whether it's a minimum,maximum or inflexion point) which depends on the sign of the scale factor second derivative along these curves and thus on $w$. The contact points of the parametric curves (\ref{div}) with the unit circle are located in 
\begin{equation}\label{curDivalpha}
\alpha_{c\pm}=\frac{1}{2}(-1\pm\sqrt{1+4p})
\end{equation}
and plotted on figure \ref{fig4}
These contact points correspond to some values $(\omega_m,\omega_r)$ that are
\begin{eqnarray}\label{curDiv}
(\omega_{mc\pm},\omega_{rc\pm})=(\frac{2}{\sqrt{4+(-1\pm\sqrt{1+4 p})^2}},\nonumber\\
\frac{-1\pm\sqrt{1+4 p}}{\sqrt{4+(-1\pm\sqrt{1+4 p})^2}})\\\nonumber
\end{eqnarray}
and symmetrically with respect to the origin. A bouncing Universe implies an accelerated expansion at the points defined by the parametric curves (\ref{div}) where extrema of the metric function occur. From the expression (\ref{H2pp}) for $H^2$, we get for $\ddot a$
\begin{eqnarray}
\frac{2}{H^2}\frac{\ddot a}{a}=\frac{2}{H^2}(\dot H+H^2)=\nonumber\\
-\frac{p (1+3 w) \Omega_m^2}{\Omega_r}+3 w (\Omega_m+\Omega_r-1)-\Omega_r-1\label{accl}\\\nonumber
\end{eqnarray}
Hence, Universe expansion is accelerating when this last expression is positive. Replacing $\Omega_r$ by $\alpha \Omega_m$ and considering $\Omega_m\rightarrow \pm\infty$ to get the sign of this last expression when the metric function has an extremum, we deduce that it is a minimum when 
\begin{equation}\label{acc}
\mp Sign[\alpha]\left[p+3pw+\alpha(\alpha-3w(1+\alpha))\right ]>0
\end{equation}
\subsection{Example} \label{s12}
\begin{figure}[h]
\centering
\includegraphics[width=8cm]{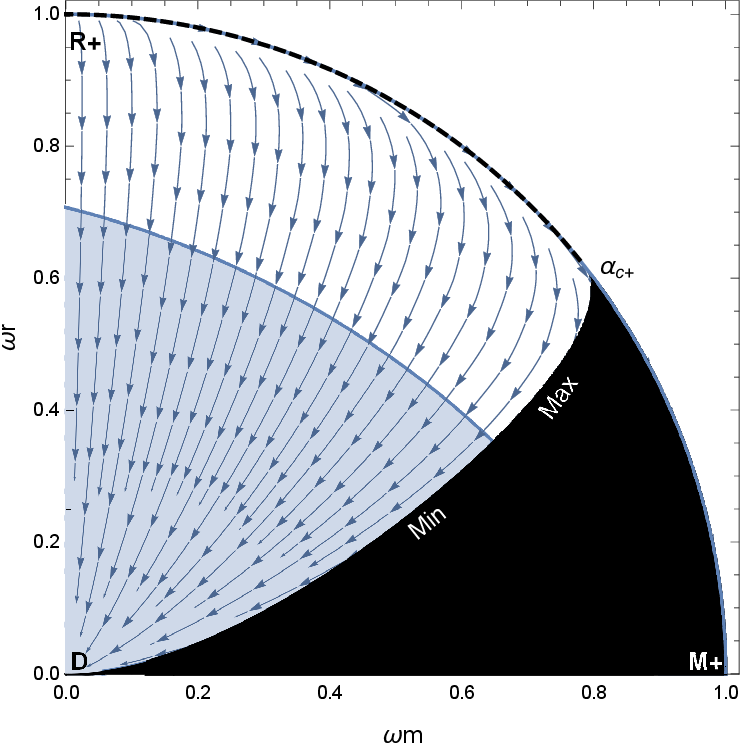}
\caption{\scriptsize{\label{fig8}Trajectories for General Relativity with a vacuum energy ($w=-1$) in the phase space $(\omega_m,\omega_r)$ when densities are positive and Universe is closed. We set $p=1.4$. The gray area represents the points in the phase space where $\ddot a>0$, indicating expansion acceleration. The minimum and maximum of the scale factor are marked along the black area outline defined by parametric curve (\ref{div}). Its contact point with the unit circle occurs in $\alpha_{c+}$ and extrema of the scale factor satisfy $0<\alpha<\alpha_{c+}$.}}
\end{figure}

To illustrate the aforementioned formalism, we plot on figure \ref{fig8} the $\Lambda CDM$ phase space where $w=-1$ and $p=1.4$ (this value is chosen for illustration purpose) with all energy densities being positive. The phase space is bounded by the unit circle and the black area outline defined by parametric curves (\ref{div}) when $\alpha>0$ and $\Omega_d>0$ (the latter inequality can be checked in the phase space using equations (\ref{Otoo}) in the constraint (\ref{cons})). The scale factor reaches an extremum on the black area outline whose contact point $\alpha_{c+}$ with the unit circle is defined by (\ref{curDivalpha}). Universe is accelerating ($\ddot a>0$) when trajectories lie within the gray area defined by equation (\ref{accl}) with $(\Omega_m,\Omega_r)$ given by (\ref{Otoo}).\\
The arrows along the trajectories indicate an increasing scale factor corresponding to an increasing $t$ time as Universe expands and a decreasing one as it contracts. The trajectories show that scale factor evolution can be monotonous, going from the source point $R_+$ to the sink point $D$. It may exhibit a maximum, originating from the $R_+$ source point during the expansion, reaching a maximum of the expansion on the black area outline and going back to the $R_+$ sink point during the contraction phase. Last, it may display a minimum, starting from the $D$ source point during the contraction phase, reaching a minimum of the expansion on the black area outline and going back to the $D$ sink point during the expansion phase. We have verified numerically that we can reproduce the phase space trajectories of figure \ref{fig8} using the original field equations (\ref{eqH2}-\ref{eqd}).
\subsection{A negative dark fluid energy density} \label{s13}
\begin{figure}[h]
\centering
\includegraphics[width=8cm]{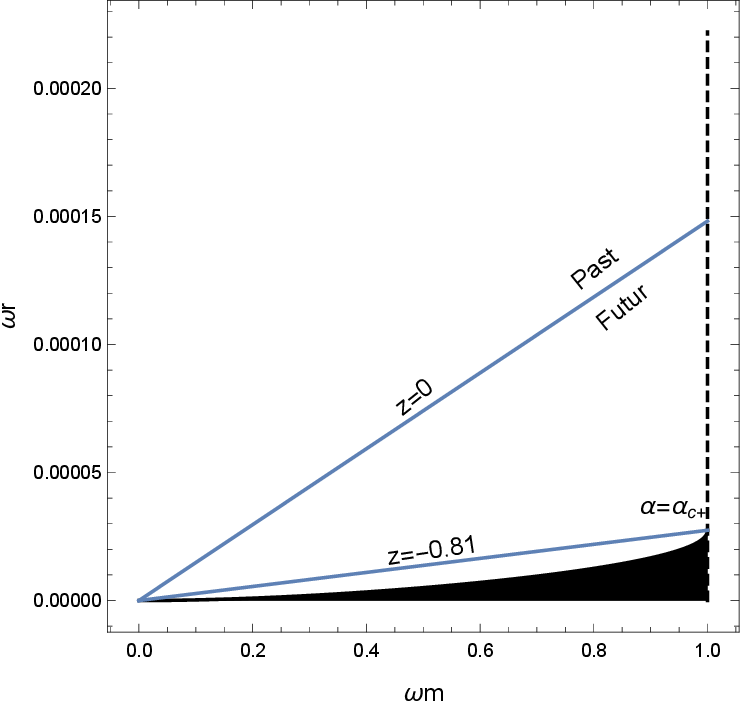}
\includegraphics[width=8cm]{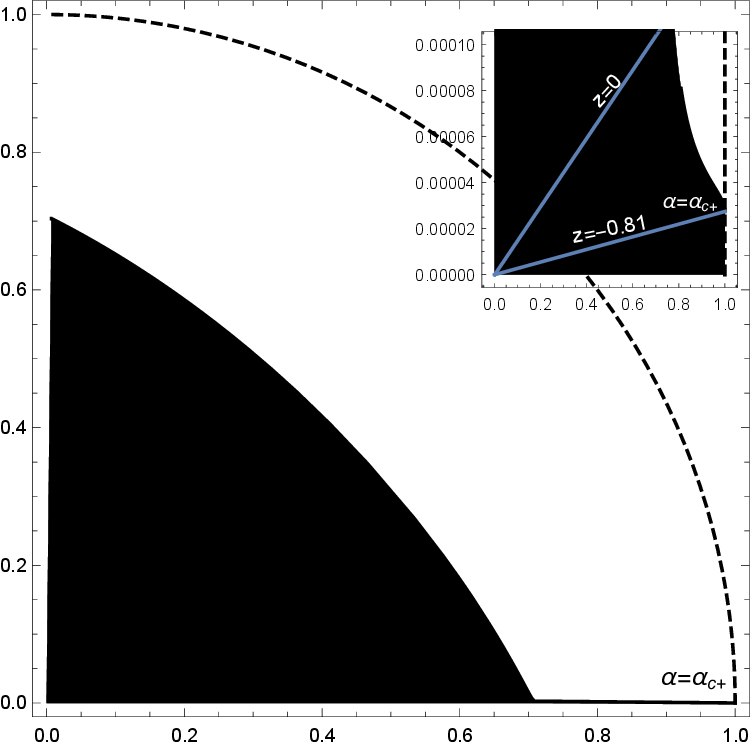}
\caption{\scriptsize{\label{fig3}Black area when $p=0.27\times 10^{-4}$ and $\Omega_d>0$ (first graph) or $\Omega_d<0$ (second graph). In the first graph we plot the lines corresponding to the redshift $z=0$ and $z=-0.81$ in agreement with the observations. It reveals that, when all the densities are positive, a bounce (or any extremum of the metric function) is possible only in the future for $z<-0.81$. In the second graph where $\Omega_d<0$, the inset shows what looks like the black area around $\omega_r=0$ and the lines for the redshift $z=-0.81$ and $z=0$ indicating that, this time, a bounce is possible when $z>-0.81$ and thus possibly in the past.}}
\end{figure}

At present time, it is assumed that the effective dark fluid density is positive, although this assumption may be subject to debate\cite{SenAdil21}. However, we are going to show that any theories describing the whole universe history that can be cast into General Relativity with an effective dark fluid have a bounce in the past in agreement with the observed Universe only if this effective dark fluid energy density is negative at the bounce. Let us note that a negative density does not necessarily imply it is unphysical\cite{ChaBar19, Far18, NemJos15}, especially when it represents a mathematical effective energy density of a model where all physical densities are positive.\\
First, we determine the range of values of the constant $p$ consistent with observations, namely $\Omega_{m(obs)}\sim 0.27$, $\Omega_{r(obs)}\sim 0.4\times 10^{-4}$ and $0<\Omega_{k(obs)}<0.05$. Error bars are omitted for $\Omega_{m(obs)}$ and $\Omega_{r(obs)}$ as our focus lies not in fitting specific cosmological models but we leave some error bars on the curvature parameter $\Omega_{k(obs)}$ since observations still hesitate between a flat or a closed Universe. Taking the initial time $N=0$ today for the rest of this paper, we thus have $\Omega_{obs}=\Omega_0$ yielding $0<p<p_{max}=0.27\times 10^{-4}$ whatever $w$. The shapes of the black areas, that only depend on $p$ and are therefore model-independent, are illustrated for the value $p=0.27\times 10^{-4}$ (i.e. $\Omega_{k(obs)}=0.05$) on figure \ref{fig3} when $\Omega_d>0$ and $\Omega_d<0$ represented respectively on the first and second graph. Furthermore, with $N=0$ denoting the present epoch, one can plot the line corresponding to the redshift $z=0$ on the first graph with $\Omega_d>0$ since then 
$$
\frac{\omega_r}{\omega_m}=\frac{\Omega_{r0}}{\Omega_{m0}}(1+z)=1.5\times 10^{-4}(1+z)
$$
The past is above this line and the future, below. This shows that any extremum of the metric function when the densities are positive can only arise in the future thanks to the positive curvature. We can then calculate the minimum redshift for an extremum that depends on the maximum value of the contact point $\alpha_{c+}$.
It is such as $\alpha_{c+}\lesssim p_{max}$ (see (\ref{curDivalpha}) when $p$ is small). The contact point is shown in figure \ref{fig3} for $p=0.27\times 10^{-4}$. Given that an extremum of the metric function always occurs for $\alpha<\alpha_{c+}$ and $\alpha=\frac{\Omega_r}{\Omega_m}=\frac{\omega_r}{\omega_m}$, it follows that an extremum of the scale factor consistent with observational values of the densities and curvature parameters could only occur in the future when $\frac{\Omega_r}{\Omega_m}<p_{max}$ or equivalently a redshift $z<-0.81$ (also plotted on figure \ref{fig3}) or even smaller when $\Omega_{k 0}<0.05$. This is no more the case if the effective dark fluid density is negative\cite{Cal21, Sah22} as shown in the second graph of figure \ref{fig3}. Here, an extremum of the metric function can occur for any redshift $z>-0.81$ as shown by the black area outline in the inset of figure \ref{fig3}. Consequently, General Relativity with matter, radiation and an effective dark fluid describes the whole Universe history with a bounce in the past and respects the aforementioned observational constraints, if the effective dark fluid density is negative at the bounce, thereby violating the weak energy condition. We underline that this result is model-independent: it does not depend on $w$ and thus on the cosmological model that it represents (dark energy, brane , $f(R)$, etc., any models that can be rewritten as General Relativity with an effective dark fluid). It's worth noting that such a violation isn't necessary for a bounce (for that, a positive curvature can be sufficient) but to have a bounce in the past in agreement with the observations. This also implies that a theory describing a bouncing Universe with positive energy densities and that thus bounces only thanks to a positive curvature cannot describe the \emph{whole} universe history and be in agreement with the observations.\\
A last important thing is that the redshift $z$ at which a dark fluid density could change sign is also determined by the observational data, still independently on $w$. Rewriting equation (\ref{eqH2}) as
$$
\Omega_d=1-\Omega_{m0}(1+z)^3-\Omega_{r0}(1+z)^4+\Omega_{\kappa 0}(1+z)^2 
$$
$\Omega_d=0$ for $0.54<z<0.61$ when $\Omega_{k(obs)}$ varies in the range $0<\Omega_{k(obs)}<0.05$. Note that it does not mean that any dark fluid density changes sign for $0.54<z<0.61$ but that if it can do it and thus be vanishing, it will arise in this redshift interval.
\section{Bouncing cosmology in the phase space}\label{s2}
In this section, we consider several cosmological models - namely a non linear dark fluid with two cosmological constants $\rho_\Lambda$ and $\rho_*$, a RS2 cosmology with a single $Z_2$ symmetric brane with a tension $\sigma$ and a cosmological constant $\Lambda_4$ and last, a $f(R)=R+mR^n+\Lambda$ theory in the Palatini formalism - and we cast them into General Relativity with matter, radiation and an effective dark fluid when necessary. Assuming that $N=0$ today, we look for the constraints on these free parameters ($\rho_\Lambda$, $\rho_*$, $\sigma$, $\Lambda_4$, $m$ and $n$) allowing a bounce at early time with a negative effective dark fluid density able to change sign in $0.54<z<0.61$.
\subsection{Non linear dark fluid with two cosmological constants}\label{s21}
\begin{figure}[h]
\centering
\includegraphics[width=8cm]{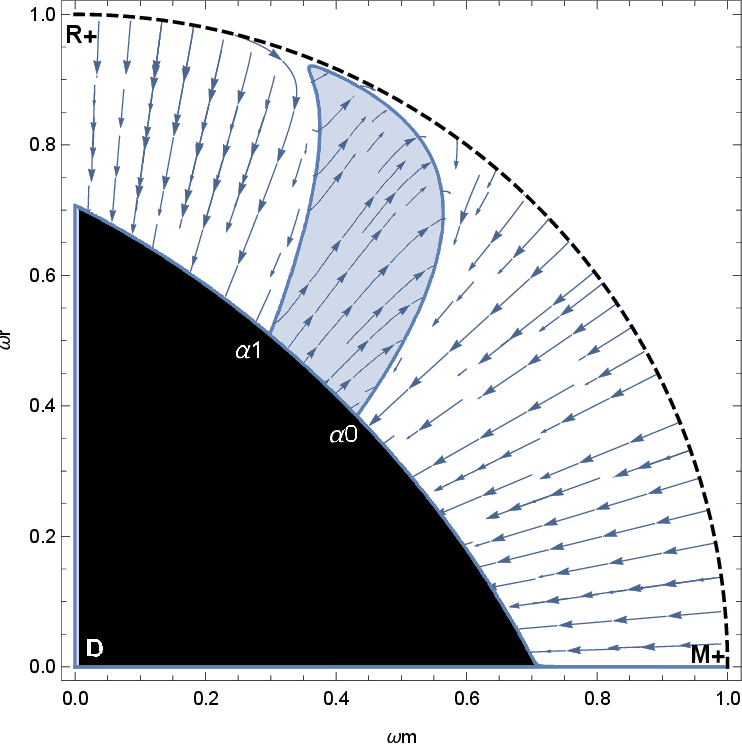}
\caption{\scriptsize{\label{fig11}Phase space for the dark fluid with two cosmological constants $\rho_L=-15$ and $\rho_s=-1.2$ with $p=0.27\times 10^{-4}$ and $\rho_d<0$. Bouncing occurs on the black area outline for $0.88<\alpha<1.6$ but the trajectories are tangent to the unit circle and $\rho_d$ never changes sign.}}
\end{figure}

In this model studied in \cite{BurBru23}, it is assumed that dark fluid conservation equation (\ref{eqd}) has two fixed points ($\dot\rho_d =0$), corresponding to two cosmological constants $\rho_\Lambda$ and $\rho_*$ that could take values in agreement with the different early and late times values of a cosmological constant, thus alleviating the cosmological constant problem. Its bouncing properties have been studied in \cite{BurBru23} for positive dark fluid energy density and we want to do the same with a negative dark fluid energy density. In \cite{BurBru23}, the following relation is assumed between the pressure $p_d$ and density $\rho_d$ of the dark fluid  
$$
p_d=-\rho_\Lambda+\frac{\rho_\Lambda}{\rho_*}\rho_d-\frac{\rho_d^2}{\rho_*}
$$
with $\rho_{\Lambda}$ and $\rho_*$, two constants. The corresponding equation of state is
$$
w=-\frac{\rho_\Lambda}{\rho_d}+\frac{\rho_\Lambda}{\rho_*}-\frac{\rho_d}{\rho_*}
$$
with 
$$
\rho_d=\pm 3\frac{H_0^2\Omega_{m0}^4}{\kappa \Omega_{r0}^3}\frac{\omega_r^3}{\omega_m^4}\sqrt{1-\omega_m^2-\omega_r^2}
$$
vanishing on the unit circle. Putting
$$
\rho_s=\rho_* \frac{\kappa \Omega_{r0}^3}{3H_0^2\Omega_{m0}^4}
$$
$$
\rho_L=\rho_\Lambda \frac{\kappa \Omega_{r0}^3}{3H_0^2\Omega_{m0}^4}
$$
$w$ rewrites as 
$$
w=\frac{\rho_L}{\rho_s}+\rho_L\frac{\omega_m^4}{\omega_r^3} \frac{1}{\sqrt{1-\omega_m^2-\omega_r^2}}+\frac{1}{\rho_s}\frac{\omega_r^3}{\omega_m^4} \sqrt{1-\omega_m^2-\omega_r^2}
$$
Along the black area outline shown on the second graph of figure \ref{fig3}, the sign of $\ddot a$ is this of
$$
acc_{black}=\frac{3\alpha^2\rho_L(\alpha^2+\alpha-p)+\alpha^2\rho_s(\alpha^2+\alpha-p)-\rho_s(2\alpha^4+\alpha^3-3\rho_L)+3\alpha^4(\alpha^2+\alpha-p)^2}{2\rho_s\alpha^2}
$$
By studying numerically this equation, we can show that there are always some initial conditions $(\omega_m,\omega_r)$ allowing a bounce:
\begin{itemize}
\item When $(\rho_L,\rho_s)$ are positive, there is a bounce whatever the phase space trajectory.
\item When $\rho_L<0$ and $\rho_s>0$, there is a bounce for any phase space trajectory intersecting the black area outline when $\alpha>\alpha_0$, with $\alpha_0$ the only positive root of $acc_{black}=0$.
\item When $\rho_L>0$ and $\rho_s<0$, there is a bounce for any phase space trajectory intersecting the black area outline when $\alpha<\alpha_0$, with $\alpha_0$ the only positive root of $acc_{black}=0$.
\item When $(\rho_L,\rho_s)$ are negative, there is a bounce for any phase space trajectory intersecting the black area outline when $\alpha_0<\alpha<\alpha_1$, with $\alpha_0$ and $\alpha_1$ the only positive roots for $acc_{black}=0$. An example of such a phase space is plotted on figure \ref{fig11} for illustration purpose.
\end{itemize}
Although a bounce is always possible whatever the signs of $(\rho_L,\rho_s)$, one can show that the sign of $\rho_d$ never changes even if it vanishes on the unit circle. By calculating $d\omega_r/d\omega_m$ on the unit circle, one shows that the trajectories always remain tangent to this circle. We conclude that if Universe can bounce thanks to a non linear dark fluid with two cosmological constants, its density will never become positive. Hence, this model could potentially describe the epoch of a bounce when $\rho_d<0$ but not beyond.
\subsection{Brane cosmology}\label{s21}
\begin{figure}[h]
\centering
\includegraphics[width=7cm]{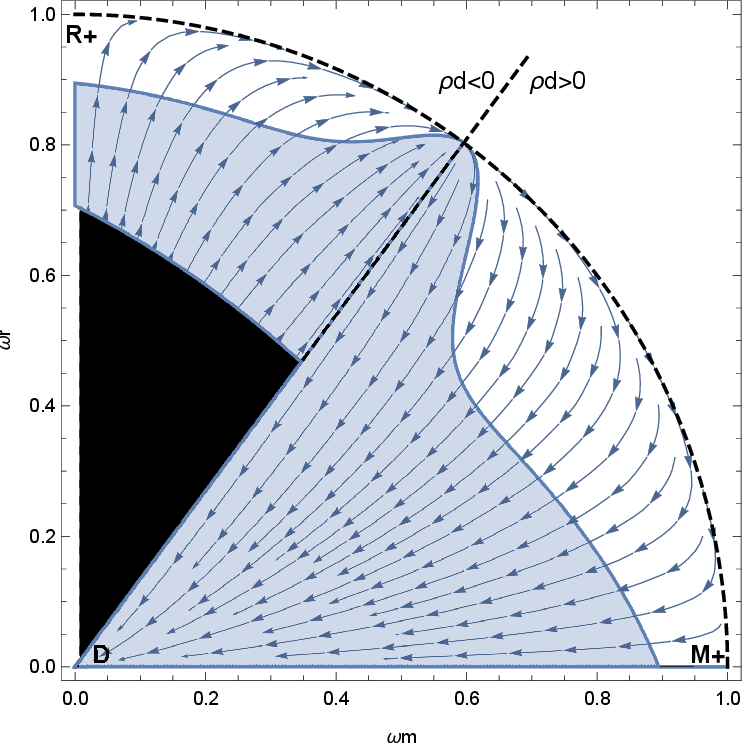}
\includegraphics[width=7cm]{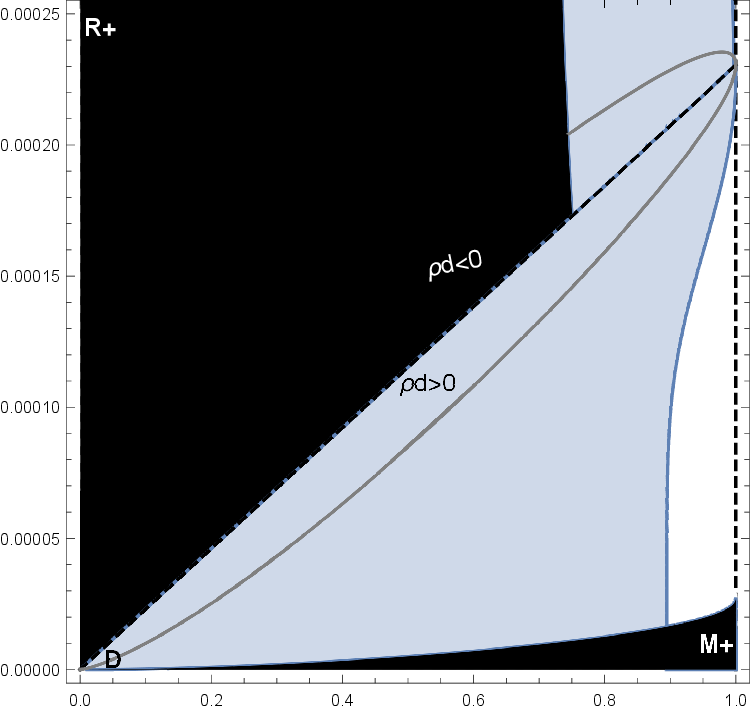}
\caption{\scriptsize{\label{fig9}First graph: Phase space for the RS2 cosmology when $(-2\Lambda\sigma)^{1/6}=1.34$ (this value is taken for illustration purpose) and $p=0.27\times 10^{-4}$. The dashed straight line separates the phase space points with $\rho_d<0$ above and $\rho_d>0$ below. The trajectories describe a Universe with a minimum of its expansion when $\rho_d<0$ and tending to a de Sitter Universe at both ends in $(\omega_m,\omega_r)=(0,0)$ with $\rho_d>0$. Second graph: The same phase space but with $(-2\Lambda\sigma)^{1/6}=2.3\times 10^{-4}$. Only one trajectory is plotted that corresponds to observational values $(\Omega_{m0},\Omega_{r0})=(0.27,4\times 10^{-4})$ today. This bounce is clearly not radiation dominated.}}
\end{figure}

We consider a RS2 cosmology with a single $Z_2$ symmetric brane \cite{RanSun99, Cli12} whose Friedmann equation is
$$
H^2=\Lambda_4+\frac{8\pi G_4}{3}\rho_m(1+\frac{\rho_m}{2\sigma})+\frac{\mu}{a^4}-\frac{k}{a^2}
$$
$\Lambda_4$ is a cosmological constant on the brane and $\sigma$ the tension. We rewrite it as General Relativity with an effective dark fluid by putting
$$
\rho_r=\rho_{r0}a^{-4}=\frac{3\mu}{8\pi G_4}a^{-4}
$$
$$
\rho_d=\frac{\rho_m^2}{2\sigma}+\frac{3\Lambda_4}{8\pi G_4}
$$
Defining also the constant
$$
\Lambda = H_0^{-4} \frac{3\Lambda_4}{8\pi G_4}\frac{\Omega_{r0}^6}{\Omega_{m0}^8}
$$
$\rho_d$ rewrites
$$
\rho_d=H_0^4\frac{\Omega_{m0}^8}{\Omega_{r0}^6}(\Lambda+\frac{1}{2\sigma}\frac{\omega_r^6}{\omega_m^6})
$$
Hence, when $\Lambda$ and $\sigma<0$, $\rho_d<0$ but does not change sign and we exclude this case. But when $\Lambda$ and $\sigma$ have opposite signs, positive and negative values of $\rho_d$ are separated in the phase space by the line $\omega_r=(-2\Lambda\sigma)^{1/6}\omega_m$. Then $\rho_d<0$ above this line if $\Lambda>0$ and below this line otherwise. On a trajectory, the sign of $\rho_d$ changes at the intersect between this line and the unit circle (where $\rho_d=0$), i.e. at the point
$$
(\omega_m,\omega_r)=(\sqrt{\frac{-(-2\Lambda\sigma)^{2/3}+(-2\Lambda\sigma)^{1/3}-1}{2\Lambda\sigma-1}},2^{1/6}(-\Lambda\sigma)^{1/6}\sqrt{\frac{-(-2\Lambda\sigma)^{2/3}+(-2\Lambda\sigma)^{1/3}-1}{2\Lambda\sigma-1}})
$$ 
We also derive the effective dark fluid equation of state allowing to close the dynamical system (\ref{eq1}-\ref{eq2})
$$
w=\frac{\omega_r^6-2 \Lambda  \sigma  \omega_m^6}{2 \Lambda  \sigma  \omega_m^6+\omega_r^6}
$$
For this theory, we can find analytically the equilibrium points stability: $R$ and $M$ are saddle whereas $D$ is a source above the line $\omega_r=(-2\Lambda\sigma)^{1/6}\omega_m$ and a sink otherwise during the expansion phase (and the opposite during the contraction phase). Along the black area outline, the sign of $\ddot a$ is this of
$$
\frac{2 \alpha^8+3 \alpha^7-4 \alpha^6 p-8 \alpha^2 \Lambda  \sigma -6 \alpha \Lambda  \sigma +4 \Lambda  p \sigma }{\alpha^6+2 \Lambda  \sigma }
$$
We then derive that when $\Lambda<0$ and $\sigma>0$, Universe expansion has a maximum. There is a minimum with $\rho_d$ changing its sign only when $\Lambda>0$ and $\sigma<0$ meaning that the cosmological constant $\Lambda_4$ on the brane is positive and the tension $\sigma$ negative\cite{Bur22}. An example of such a phase space is shown on the first graph of figure \ref{fig9}. Moreover, this minimum is such as $\alpha>(-2\Lambda\sigma)^{1/6}$ and there are thus some trajectories for which the bounce is dominated by radiation.\\
Hence, at this stage, we know that a bounce can exist at early time with $\rho_d<0$, and that the dark fluid density can change sign. We now determine the conditions such as the redshift for which $\rho_d$ changes sign is in agreement with observations, i.e. in the range $0.54<z<0.61$ (see subsection \ref{s13}). Since $\frac{\omega_r}{\omega_m}=\frac{\Omega_{r0}}{\Omega_{m0}}(1+z)$ and $\rho_d$ have to change sign in $\frac{\omega_r}{\omega_m}=(-2\Lambda\sigma)^{1/6}$, we get from observations that these inequalities are respected if $2.28\times 10^{-4}<(-2\Lambda\sigma)^{1/6}<2.38\times 10^{-4}$. This means that the line $\omega_r=(-2\Lambda\sigma)^{1/6}\omega_m$ separating the points with $\rho_d<0$ and $\rho_d>0$ should be very closed from the $\omega_r=0$ line as shown in the second graph of figure \ref{fig9}. In these conditions, trajectories with a radiation dominated bounce (and thus at early time) should also have a very small $\Omega_{r0}$ today since they pass closer from the $\omega_r=0$ line than other trajectories with matter dominated bounce. So, is the observational value $\Omega_{r0}=4\times 10^{-4}$ small enough for a radiation dominated bounce ? Indeed, considering the trajectory in agreement with the data $(\Omega_{m0},\Omega_{r0})=(0.27,4\times 10^{-4})$ or equivalently $(\omega_{m0},\omega_{r0})=(0.32,4.8\times 10^{-5})$, the second graph of figure \ref{fig9} shows that the corresponding bounce is matter dominated. Hence, the RS2 cosmology, even with a negative dark fluid able to change its sign, cannot lead to an early time bounce dominated by radiation and be in agreement with the present day observations.
\subsection{$f(R)=R+mR^n+\Lambda$}\label{s22}
\begin{figure}[h]
\centering
\includegraphics[width=8cm]{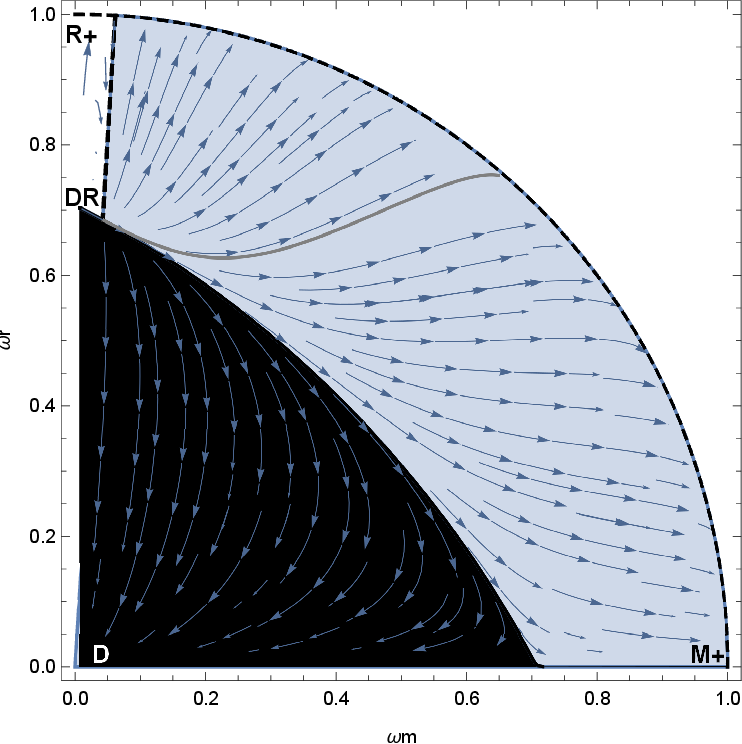}
\caption{\scriptsize{\label{fig10}Phase space for $n=2.03$ and $m=0.78$ (we do not chose $n$ closer from $2$ otherwise the bounce is not visible on the figure, but the closer $n$ from $2$, the larger the value of $\alpha$ at the bounce). A choice of initial conditions select the gray trajectory as respecting equation (\ref{Hfr}) at an initial time $N=0$. The bounce occurs at the intersection between the black area outline and the curve (\ref{wDiv}) separating the positive and negative values of $\rho_d$ (dashed black line).}}
\end{figure}

We now consider the $f(R)=R+mR^n+\Lambda$ theory in the Palatini formalism with $n>0$. We first assume that near the bounce, the $mR^n$ term is dominating and look for the values of $n$ in agreement with a dominated radiation bounce with a negative effective dark fluid energy density. $f(R)$ theories write in general
$$
H^2=\frac{2k(\rho_m+\rho_m+f_R R-f)}{6f_R\xi}-\frac{k}{a^2\xi}
$$
with $f_R=df/dR$ and
$$
\xi=(1-\frac{3}{2}\frac{(Rf_R-2f)f_{RR}}{f_R(Rf_{RR}-f_R)})^2
$$
\begin{equation}\label{consR}
f_R R-2f=-\kappa\rho_m
\end{equation}
This last equation gives $0=-\kappa\rho_m$ when $n=2$ meaning that it cannot be satisfied in presence of matter. We discard this special value for $n$. We put 
\begin{equation}\label{mmb}
m=\frac{\bar m}{2-n}(\frac{\Omega_{r0}^3}{H_0^2\Omega_{m0}^4})^{n-1}
\end{equation}
Then, using 
$$
\frac{d(H^2)}{dN}=-3\Omega_m-4\Omega_r-3(1+w)\Omega_d+2\Omega_k
$$
it comes for the effective dark fluid equation of state
\begin{eqnarray}\label{wr}
w&=&\{6 n^2 p \omega_r^5 \bar m^{1/n} \omega_m^{3/n} (-\sqrt{-\omega_m^2-\omega_r^2+1}+\omega_m+2 \omega_r)+3^{1/n} \omega_m \omega_r^{\frac{3}{n}+2} \{3 (n^2-4 n+3) \omega_m^2 \omega_r-\nonumber\\
&&(n-3) p \omega_m^2 [(2 n-3) \omega_m+2 (n-2) \omega_r]-(n-1) \omega_m \omega_r [3 (n-3) \sqrt{-\omega_m^2-\omega_r^2+1}+\nonumber\\
&&(21-8 n) \omega_r]+2 (n-2) \omega_r^2 [(3-2 n) \sqrt{-\omega_m^2-\omega_r^2+1}+3 (n-1) \omega_r]\}\}/\nonumber\\
&&\{3 n\omega_r^3 \sqrt{-\omega_m^2-\omega_r^2+1} [6 n p \omega_r^2 \bar m^{1/n} \omega_m^{3/n}+3^{1/n} \omega_m \omega_r^{3/n} ((n-3) \omega_m+2 (n-2) \omega_r)]\}
\end{eqnarray}
The term $\bar m^{1/n}$ appearing in $w$ means that either $\bar m>0$ or $\bar m<0$ with $n$ being the inverse of an integer (and then $0<n<1$). This is the replacement (\ref{mmb}) of $m$ by $\bar m$ that allows to eliminate $(H_0,\Omega_{m0},\Omega_{r0})$ from the expression for $w$. It has another consequence. It also allows to rewrite the Hubble function as 
\begin{equation}\label{Hfr}
H^2=-\frac{2 H_0^2 n \Omega_{m0}^4 \bar m^{-1/n} \omega_m^{-\frac{3}{n}-2} \left(6 n p \omega_r^2 \bar m^{1/n} \omega_m^{3/n}+3^{1/n} \omega_m \omega_r^{3/n} ((n-3) \omega_m+2 (n-2) \omega_r)\right)}{3 (n-3)^2 \Omega_{r0}^3}
\end{equation}
Hence, if we choose $\bar m$, $n$ and an initial condition in $\omega_m=\omega_{m0}$, this last equation determines $\omega_{r0}$ since $H_0^2$ disappears when $H=H_0$. This means that in the phase space $(\omega_m,\omega_r)$, only one trajectory respects the constraint on the initial conditions imposed by (\ref{Hfr}).\\
Replacing $w$ in (\ref{var1}-\ref{var2}), one derives that there is an equilibrium points DR in $(\omega_m,\omega_r)=(0,1/\sqrt{2})$ corresponding to a singularity. Moreover, $w$ diverges and $\rho_d$ changes sign on the unit circle and on the curve
\begin{equation}\label{wDiv}
6 n p \omega_r^2 \bar m^{1/n} \omega_m^{3/n}+3^{1/n} \omega_m \omega_r^{3/n} ((n-3) \omega_m+2 (n-2) \omega_r)=0
\end{equation}
This separates the phase space points in several parts related to the sign of $\rho_d$. On the black area outline, the sign of $\ddot a$ is this of
\begin{equation}\label{RnAcc}
\frac{3^{1/n} (n-3) \alpha^{3/n} [2 \alpha  (n-2)+2 n-3]}{2 n [6 \alpha ^2 n p \bar m^{1/n}+3^{1/n} (n-3) \alpha^{3/n}+2\ 3^{1/n} (n-2) \alpha^{\frac{n+3}{n}}]}
\end{equation}
and the Hubble function (\ref{Hfr}) becomes 
\begin{equation}\label{RnCons}
\frac{2H_0^2\bar m^{-1/n} n [-6 \alpha ^2 n p \bar m^{1/n}-3^{1/n} (n-3) \alpha^{3/n}-2\ 3^{1/n} (n-2) \alpha^{\frac{n+3}{n}}]\Omega_{m0}^4}{3(n-3)^2\Omega_{r0}^3}=0
\end{equation}
since then $H=0$. This last expression vanishes if there are positive and negative terms in its numerator that is not possible if $n\geq 3$, meaning that $0<n< 3$. Moreover it vanishes when $H^2$ tends to $0^+$ and one remarks that the numerator of (\ref{RnCons}) has the opposite sign of the denominator of (\ref{RnAcc}) for $\ddot a$. Hence, since $n< 3$, equations (\ref{RnCons}) tends to $0^+$ when we have a minimum ($\ddot a>0$) only if the numerator of $\ddot a$ is negative i.e.
\begin{equation}\label{accAlCons}
2 \alpha  (n-2)+2 n-3>0
\end{equation}
We deduce from this last inequality, that there is no acceleration (and thus no bounce) on the black area outline when $n\leq 3/2$. Thus, a minimum could exist only when $3/2<n<3$ and consequently with $\bar m>0$. We now examine the conditions for its existence when radiation is dominating ($\alpha>>1$). 
\begin{itemize}
\item When $3/2<n<2$, $m>0$. If a minimum occurs for $\alpha>>1$ (typically $\alpha>10 000$ for a bounce at a larger redshift than this of the nucleosynthesis around $10^8$), the condition (\ref{accAlCons}) is satisfied if $2 \alpha  (n-2)\leq O(1)$ and thus $n\approx 2^-$. Then, (\ref{RnCons}) vanishes in
$$
\alpha \pm\approx\frac{12 \bar m p^2+n-2\pm 2 \sqrt{6} \sqrt{\bar m p^2 \left(6 \bar m p^2+n-2\right)}}{2 (n-2)^2}
$$
with $n\approx 2^-$. $\alpha \pm$ also correspond to the intersection of (\ref{wDiv}) ($\rho_d=0$) with the black area outline. Note that there are two solutions $\alpha \pm$ but only one trajectory in the phase space in agreement with (\ref{Hfr}).
\item When $2<n<3$, $m<0$. The condition (\ref{accAlCons}) is satisfied. In (\ref{RnCons}) since $\alpha^{3/n}<<\alpha^{\frac{n+3}{n}}$ the numerator is negative and not vanishing but if $n\approx 2^+$. Then, once again, their is a bounce in $\alpha_\pm$. This case is illustrated on figure \ref{fig10}.
\end{itemize}
It follows that for any $f(R)$ theories tending to $mR^n$, a bounce can occur when $3/2<n<3$ and during the radiation dominated era (i.e. at early time) only when $n\approx 2$.\\
We now consider the theory $f(R)=R+mR^n+\Lambda$, able to describe the whole Universe history, with $n\approx 2$ and $\Lambda$ the cosmological constant. We look when the corresponding effective dark fluid density changes sign. We thus calculate $\Omega_d$ (a long mathematical expression that we do not reproduce here) and look for the values of $\bar m$ such as $\Omega_d=0$ when $0.54<z<0.61$ and for various values of $n$ such as $\mid n-2\mid<10^{-3}$. We always find that $\bar m>10^4>>1$. When plotting the corresponding solutions, one always find that $\omega_r<<1$ at any time (the trajectories are always close to the $\omega_m$ axe) and thus in particular during the bounce that is then never radiation dominated. To understand why, we calculate $H^2$ during the matter radiation equality $\omega_m=\omega_r$ and find for $\bar m>>1$ that 
$$
H^2\approx -\frac{16 p \Omega_{m0}^4 (3 H_0^3 \Omega_{m0}^4-2 H_0 \Lambda  \Omega_{r0}^3)^2}{\Omega_{r0}^3 (3 H_0^2 \Omega_{m0}^4-8 \Lambda  \Omega_{r0}^3)^2}<0
$$
Therefore, when $\bar m$ is large as mandated by observations for the transition from negative to positive dark fluid density, there is no matter-radiation equality that is not in agreement with our Universe. We conclude that the $f(R)=R+mR^n+\Lambda$ theory with $n\approx 2$ as required to have a radiation dominated bounce cannot describe our whole Universe history.
\section{Conclusion}\label{s3}
We have established a dynamical system allowing to study a cosmological bounce for any gravitation theory that can be cast into General Relativity with matter, radiation, an effective (or not if one studies a General Relativity model as in section \ref{s21}) dark fluid and curvature. This system is parameterized by a constant $p$ which can be constrained by observations $\Omega_{m(obs)}\sim 0.27$, $\Omega_{r(obs)}\sim 0.4\times 10^{-4}$ and $0<\Omega_{k(obs)}<0.05$ leading to $0<p<0.27\times 10^{-4}$. This allows to reduce the dynamical system to a 2D system with normalised variables $(\omega_m,\omega_r)$ related to matter and radiation. Then, the points of the phase space corresponding to extrema of the metric function (both minimum and maximum) are also determined by the value of $p$ and are independent from the cosmological model or equivalently, the dark fluid equation of state.\\
A first result indicates that if the effective dark fluid density is positive at the bounce, observations show that a bounce is only possible in the future for $z<-0.81$. Consequently, the effective dark fluid density must be negative at the bounce that can then occur for $z>-0.81$, i.e. possibly in the past. Thus, while a positive curvature (closed Universe) might be sufficient to yield a bouncing cosmology, it is insufficient to achieve a bouncing cosmology consistent with observations when considering that the \emph{whole} Universe history is described by a cosmological model bridging early and late epochs and allowing to late time observations (in $z=0$) to constraint an early time bounce. With this in mind, we proceed to investigate three cosmological models that either belong to or can be reformulated within General Relativity with an effective dark fluid. Assuming that the effective dark fluid density is negative at the bounce, we determined the conditions for the existence of a bounce at early time (i.e. when radiation dominates over matter) and whether the effective dark fluid density sign changes in the range $0.54<z<0.61$ required by observations.\\
The first model we considered is a non linear dark fluid with two cosmological constants $\rho_\Lambda$ and $\rho_*$. These two constants are used\cite{BurBru23} to solve the cosmological constant problem and this model is thus built to describe continuously Universe from early to late time. Although it is always possible to have a bounce whatever $\rho_\Lambda$ and $\rho_*$, the dark fluid never changes sign and thus this model fails to describe a whole Universe history with a bounce in agreement with observations. The second model we considered	 is a RS2 cosmology with a single $Z_2$ symmetric brane. There is a minimum when the cosmological constant $\Lambda_4$ on the brane is positive and the tension $\sigma$ is negative. This minimum occurs when $\alpha>(-2\Lambda\sigma)^{1/6}$ or $a_{min}<\frac{\Omega_{r0}}{\Omega_{m0}}(-2\Lambda\sigma)^{-1/6}$ and can thus be dominated by radiation. However, we also calculate that the dark fluid density changes sign in $0.54<z<0.61$ only if $2.28\times 10^{-4}<(-2\Lambda\sigma)^{1/6}<2.38\times 10^{-4}$ and unfortunately this last constraint implies that the bounce is never radiation dominated and cannot arise at early time. The last model we considered is this of the $f(R)=R+mR^n+\Lambda$ theory in the Palatini formalism. A radiation dominated bounce needs that $n\approx 2$ but observations imply then that $\bar m>>1$ such as the effective $\rho_d$ changes sign in $0.54<z<0.61$ that is in disagreement with a radiation matter transition. Consequently, the $f(R)=R+mR^n+\Lambda$ theory cannot describe our whole Universe history with a radiation dominated bounce.\\
Hence, observations require a negative effective dark fluid density for a bounce at early times which then appears to be more readily achievable than with a positive $\rho_d$. However, the requirement for the bounce to occur at early time and to the effective dark fluid density to change sign (if assumed positive today) within a narrow range of redshifts generally poses a challenge for aligning such an early time bounce with present day observations whatever the cosmological model (brane, f(R), etc, anything that can be rewritten as General Relativity with an effective dark fluid).
\section{Data availability}
No new data were generated or analysed in support of this research.
\bibliographystyle{unsrt}

\end{document}